%% file: neurips_2025.tex
\documentclass{article}

\usepackage[numbers,sort&compress]{natbib}
\usepackage[dblblindworkshop, final]{neurips_2025}
\usepackage{enumitem}

\usepackage[utf8]{inputenc} 
\usepackage[T1]{fontenc}    
\usepackage{hyperref}       
\usepackage{url}            
\usepackage{booktabs}       
\usepackage{amsfonts}       
\usepackage{nicefrac}       
\usepackage{microtype}      
\usepackage{xcolor}         
\usepackage{graphicx}

\title{On the Regulatory Potential of User Interfaces for AI Agent Governance}

\author{%
  \textbf{K. J. Kevin Feng}$^{\omega}$\thanks{Equal contribution.} \and 
  \textbf{Tae Soo Kim}$^{\kappa*}$ \and
  \textbf{Rock Yuren Pang}$^{\omega}$\footnotemark[1] \and
  \textbf{Faria Huq}$^{\Gamma}$ \and
  \textbf{Tal August}$^{\Psi}$ \and
  \textbf{Amy X. Zhang}$^{\omega}$ \and
  $^{\omega}$University of Washington \quad 
  $^{\kappa}$KAIST \quad 
  $^{\Gamma}$Carnegie Mellon University \quad 
  $^{\Psi}$UIUC\\
  \texttt{kjfeng@uw.edu}
}

\begin{document}
\maketitle

\begin{abstract}
  AI agents that take actions in their environment autonomously over extended time horizons require robust governance interventions to curb their potentially consequential risks. Prior proposals for governing AI agents primarily target system-level safeguards (e.g., prompt injection monitors) or agent infrastructure (e.g., agent IDs). In this work, we explore a complementary approach: regulating \textit{user interfaces} of AI agents as a way of enforcing transparency and behavioral requirements that then demand changes at the system and/or infrastructure levels. Specifically, we analyze 22 existing agentic systems to identify UI elements that play key roles in human-agent interaction and communication. We then synthesize those elements into six high-level interaction design patterns that hold regulatory potential (e.g., requiring agent memory to be editable). We conclude with policy recommendations based on our analysis. Our work exposes a new surface for regulatory action that supplements previous proposals for practical AI agent governance.
\end{abstract}

\section{Introduction}

AI agents---compound AI systems that take actions in their environment on behalf of a user under limited direct supervision---are increasingly being built and deployed into the real world \cite{chan2025infrastructure, shavit2023practices, kapoor2024ai, feng2025levels, kolt2025governing, lazar2025governing, gabriel2025agents, chan2023harms}. Modern agents have demonstrated proficiency in autonomously pursuing complex, multi-step goals in economically-valuable domains including software engineering \cite{yang2024swe, jimenez2023swe}, online shopping \cite{yao2022webshop}, and machine learning research \cite{chan2024mle, starace2025paperbench}. Their proficiency in autonomous task completion is rapidly improving \cite{metr-long-tasks}, a source of both excitement and concern. Agents may unlock new levels of productivity and economic growth, but their deployment is accompanied by heightened risks \cite{shavit2023practices, casper2025ai}. These risks come from the inherent difficulty in anticipating these risks \cite{chan2023harms}, increased attack surfaces for malicious actors \cite{evtimov2025wasp, kumar2024refusal, patlan2025context}, accelerating the gradual disempowerment of humans \cite{kulveit2025position}, and a general loss of human control over AI systems \cite{international-safety-report}. 

Prior work in agent governance proposed ways to mitigate risks, primarily through two avenues: \textit{system-level safeguards} and \textit{agent infrastructure} \cite{chan2025infrastructure, shavit2023practices, chan2024ids, chan2024visibility, south2025authenticated, operator-sys-card, oai-agent-sys-card}. System-level safeguards fortify the AI model or agent scaffolding; it includes techniques such as training a model to seek user confirmation for sensitive actions or proactively refuse harmful requests, and building monitors for prompt injection attacks \cite{operator-sys-card, oai-agent-sys-card}. Agent infrastructure is external to the system but still mediates and influences agents' interactions with their environments; these include agent IDs \cite{chan2024ids, autogen-ids}, isolated channels for agent activities \cite{chan2025infrastructure}, and frameworks for authenticated delegation of authority \cite{south2025authenticated}. 
While promising, these approaches operate ``behind the scenes''---in model training or system architecture, limiting end users' visibility into and control over agent behavior during actual deployment.

In this work, we address this challenge by proposing a new target for regulation in user-facing agentic systems: \textit{the user interface} (UI). Regulation of UIs and UI elements is a well-established practice outside of AI for ensuring consumer safety. In the US, UK, EU, and Canada, marketing emails are legally required to contain a button for one-click unsubscribe \cite{us-unsub, uk-unsub, canada-unsub}. The Americans with Disabilities Act (ADA) and the EU Accessibility Act require UIs to be perceivable and operable by people with visual and motor impairments \cite{us-ada, eu-a11y}. The General Data Protection Regulation (GDPR) in the EU and the California Consumer Privacy Act (CCPA) both mandate the presentation of UIs for privacy consent management and opt-out controls \cite{gdpr, ccpa}. Because UI elements map to underlying system functionality and act as a key layer for user-system communication and control, regulating UIs of agents can jumpstart the implementation of certain system- and infrastructure-level interventions. An illustration of the conceptual differences between system-, infrastructure-, and UI-level interventions can be found in Appendix \ref{a:interventions}.

We begin our investigation of the regulatory potential by analyzing 22 deployed agentic systems to identify UI elements in human-agent communication. We synthesize our analysis into six high-level design patterns that can serve as targets for regulation, mapping each pattern to previously proposed system- and infrastructure-level governance for agents. We conclude with policy recommendations and fruitful avenues of collaboration for the technical, policy, and design communities.

\section{Method}
\label{s:method}
We collected and analyzed 22 agentic systems from academic papers in human-AI interaction, product releases, and open-source projects. To be considered ``agentic,'' the system needed to satisfy the following 4 inclusion criteria at the time of analysis: 1) Is publicly available; 2) Is an interactive software system; 3) Operates using multi-step workflows; and 4) Calls tools and/or executes actions. These criteria were inspired by previous literature (e.g., \cite{chan2023harms, casper2025ai, feng2025levels}), and partially for practical reasons (e.g., we cannot analyze a system in a rigorous, reproducible way if its details are not publicly available). Full definitions for our inclusion criteria can be found in Appendix \ref{a:inclusion-criteria}.

Three authors divided up the analysis using visual thematic analysis, a common method in HCI for identifying design patterns in UIs \cite{van2000interaction, feng2023ux}. Further details about our analysis process are in Appendix \ref{a:method-details}. Examples from our analysis and the full list of agentic systems are also in the Appendix.

\section{Six Design Patterns with Regulatory Potential}
\label{s:patterns}

Our analysis yielded six design patterns (see visual examples in Appendix \ref{a:pattern-examples}) that can serve as levers for regulatory action. These are not exhaustive but a starting point---new patterns may emerge as more agentic systems are deployed and UIs evolve. Fortunately, strong incentives already exist for adopting these patterns due to usability benefits. However, we still consider it important for these patterns to be regulated to prevent developers from removing them out of convenience, market pressures, or A/B testing. 

\subsection{Visible thoughts, plans, and actions}

\textbf{Description.} The agent’s reasoning, planning, and actions are represented as a step-by-step sequence that users can trace in real time or retrospectively. This trace may appear inline with the chat, alongside the cursor, or alongside components on which the agent is acting. By surfacing both what the agent is doing and why, these interfaces make the decision-making process explicit.

\textbf{Regulatory promises and challenges.} Revealing the agent’s reasoning and actions enhances transparency, accountability, and user oversight. It enables users to identify unsafe behavior, intervene, and calibrate trust in the system. However, these benefits depend on the faithfulness of the expressed reasoning to the agent’s actual internal processes \cite{emmons2025chain, chen2025reasoning}. Another challenge is managing granularity: too little detail undermines oversight, while too much risks overwhelming users and reducing usability.

\textbf{Connections with prior governance proposals.} Regulating this design pattern advances prior calls for agent oversight. Visible thought and action traces serve as an \textit{oversight layer} in the agent infrastructure, as proposed by Chan et al. \cite{chan2025infrastructure}, and operationalizes Shavit et al.'s \cite{shavit2023practices} proposed practice of \textit{legibility of agent activity} to ensure the safety of agentic AI systems.

\subsection{Mechanisms for control transfer}

\textbf{Description.} Explicit transfer of control encompasses two complementary mechanisms: interruption and takeover. Interruption allows the user to pause or stop the agent’s ongoing activities at any point. Takeover goes further by allowing the user to directly assume control over the task environment (e.g., OpenAI Operator, Orca \cite{jiang2025orca}) or steps in the agent’s workflow (e.g., Cocoa \cite{feng2024cocoa}).

\textbf{Regulatory promises and challenges.} Interruption and takeover controls are central to user agency and safety, allowing users to intervene directly in an agent’s activity. Their effectiveness depends not only on whether the agent halts cleanly when interrupted, but also on how the system manages actions already in progress or recently completed—for example, whether it should abort immediately, finish the current step, or how to design safe rollback and follow-up procedures after a takeover.

\textbf{Connections with prior governance proposals.} Control transfer implements prior calls for oversight infrastructure~\cite{chan2025infrastructure} and practices for interruptibility~\cite{shavit2023practices}. Regulating this pattern can also address the concerns raised by Kolt regarding delegation and loyalty \cite{kolt2025governing}, by allowing users to reassert authority on their interest and agency, rather than relying on the agent to decide when to involve the human.

\subsection{Watch mode}

\textbf{Description.} While operating in environments with sensitive information (e.g., financial websites or email clients), agents requires the user's direct supervision and extracts limited information from the environment. For example, while OpenAI Operator and ChatGPT agent takes regular screenshots during its operational trajectory, it pauses this behavior when it enters watch mode. In Gumbo \cite{shaikh2025creating}, the system filters out sensitive information from users' screen activities before using them as context.

\textbf{Regulatory promises and challenges.} ``Watch mode`` addresses privacy and security concerns by preventing agents from inadvertently capturing or handling sensitive information. There remains a tradeoff on the frequency of ``watch mode''. On one hand, detecting sensitive information must be robust to avoid false negatives. On the other hand, if triggered too often, it interrupts workflows with excessive control transfers. More, requiring user simply ``watching'' does not guarantee meaningful attention or control --- users may become complacent or distracted. 

\textbf{Connection with prior governance proposals.} This pattern directly implements authenticated delegation frameworks by ensuring human oversight remains active during sensitive operations~\cite{south2025authenticated}. Smiliar to the proposal to govern Tesla's Full Self-Driving (FSD) mode where drivers are required to keep their hands on the wheel but may still become inattentive. This is important, especially for high-stakes tasks such as activities involving sensitive information (e.g., banking or health).

\subsection{Customizable rule-based governance}

\textbf{Description.} The user can modify the agent's default behavior by specifying custom rules. This can include rules for how to perform consequential actions and conditions under which the agent seeks user approval. 
User-specified rules override any ``self-approval'' settings where the agent approves its own actions (e.g., Cursor's auto-run \cite{cursor-auto-run}). 

\textbf{Regulatory promises and challenges.} Agents must robustly validate and follow user-specified rules, particularly when conditions are underspecified or ambiguous. For approval requests, agents might subtly nudge users toward approval with techniques such as  reward hacking~\cite{sharma2023towards},  undermining the protective intent of these mechanisms. Additionally, malicious rules (e.g., ones that disable the agent's built-in safeguards) will need to be detected and removed.

\textbf{Connection with prior governance proposals.} This pattern can operationalize multiple governance principles including oversight layers~\cite{chan2025infrastructure}, task and resource scoping~\cite{south2025authenticated}, constraining action-space and requiring approval~\cite{shavit2023practices}, setting default behaviors~\cite{shavit2023practices}, and frameworks for delegated authority~\cite{kolt2025governing}. Generally, by allowing users to specify custom rules through the agent's UI, regulators can improve the flexibility of one-size-fits-all governance approaches imposed by developers.

\subsection{Inspectable and editable agent memory}
\textbf{Description.} Users can inspect and edit (i.e, modify, delete, add to) the agent's memory, which often includes preferences and user characteristics automatically inferred by the agent throughout the course of one or more interactions with the user. Further, options for inspecting and editing agent memory should be easily accessible and discoverable by the user. See examples from Cursor and Gumbo \cite{shaikh2025creating} in Appendix \ref{a:memory}.

\textbf{Regulatory promises and challenges.} Ensuring that users can easily access and edit an agent's memory is important for transparency, privacy, and agency. Memory may contain private information the user would not want the agent to consider in its outputs, or worse, accidentally exposed through adversarial attacks. The memory may also contain inaccurate inferences that users  should be able to correct or delete. Regulations should further ensure that developers do not disincentivize users from taking these actions by hiding them deeply in a settings menu, which be done by requiring a minimum number of clicks to access them. 

\textbf{Connections with prior governance proposals.} Kolt \cite{kolt2025governing} raised the issue of information asymmetry in agent governance, where agents having access to information that humans do not place us in a vulnerable position. Transparently exposing agents' memories and allowing users to edit them can help alleviate this problem by aligning human and agent information sources. 

\subsection{Sandboxes for agents with low-level environmental control}
\textbf{Description.} In systems that expose low-level control of the operational environment to the agent (e.g., access to the terminal on a computer), the UI clearly shows the sandboxed nature of agent activity, \textit{as well as information about the sandbox health}. Many systems display the former (see Appendix \ref{a:sandbox} for examples), but not the latter. Sandbox health may include information such as its age and whether any evidence of activity leakage has been detected.

\textbf{Regulatory promises and challenges.} The advanced capabilities of agents may render sandboxes for traditional software ineffective \cite{shavit2023practices}. Requiring sandboxes and their health to be displayed to users will encourage the development of monitoring methods that can detect when sandboxes may be broken by the agent or otherwise ineffective. Open questions include what metrics are best for tracking sandbox health, and how to test the robustness of sandboxes without incurring the risks from breakage. Poor metrics may deceive users into thinking a sandbox is healthy when it is in fact not.

\textbf{Connections with prior governance proposals.} Regulating this design pattern can help determine not only \textit{how} an agent's action space is constrained \cite{shavit2023practices}, but also help monitor \textit{the effectiveness} of that method. This is clearly already top-of-mind for developers using terminal-based coding agents, evidenced by the development of sandboxing tools like VibeKit\footnote{\url{https://www.vibekit.sh/}}.

\section{Policy Recommendations}
In light of our design patterns, we conclude with the following policy recommendations.
\begin{enumerate}
    \item \textbf{Prioritize regulation based on (lack of) existing implementation incentives.} As mentioned in Section \ref{s:patterns}, incentives for implementing some of the described design patterns already exist, as they improve system usability. Thus, regulation should first target patterns with the weakest usability incentives. For example, many developers have voluntarily made an agent's thoughts visible, but few currently communicate sandbox health. The latter should then be prioritized as a regulatory target.
    \item \textbf{Learn from lessons in dark patterns regulation.} The GDPR introduced a new wave of UI regulations to counter ``dark patterns''---UI designs that steer  users into making unintended, potentially harmful decisions for an online platform's benefit \cite{mathur2019dark}. Agent UI regulation can draw from many lessons from dark patterns regulation. For example, setting up strategic collaborations between technical, policy, and design experts should be a top priority \cite{gray2021dark}, as should empirical validation of effects on users to catch any unintended backfire effects \cite{nouwens2020dark}. 
    \item \textbf{Prepare evaluators to verify adherence to UI regulation.} National AI safety institutes (e.g., US CAISI, UK AISI) and third party evaluation organizations (e.g., METR, Apollo Research) already partner with developers for pre-deployment safety evaluations of agentic systems \cite{oai-agent-sys-card}. Policymakers should work with evaluators to build new evaluations to verify whether a UI design pattern has been implemented and behaves as expected.
    
\end{enumerate}

\bibliographystyle{plainnat}
\bibliography{ref}

\appendix
\input{appendix}

\end{document}

%% file: appendix.tex
\section{Appendix}

\subsection{Illustration of different types of agent governance interventions}
\label{a:interventions}

\begin{figure}[h]
    \centering
    \includegraphics[width=0.66\textwidth]{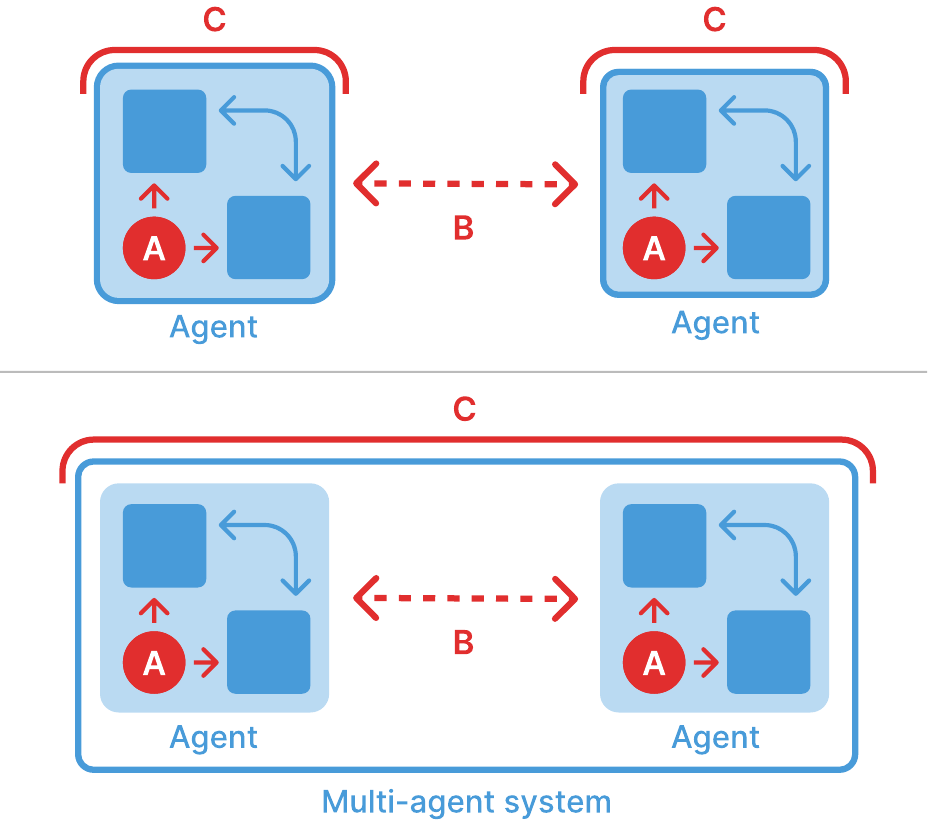}
    \caption{Governance interventions are shown in \textcolor{red}{red} for single-agent (top) and multi-agent (bottom) systems. \textbf{(A)} depicts a system-level intervention: a component (e.g., prompt injection monitor) that communicates with other components within the agent's architecture. \textbf{(B)} depicts an infrastructure-level intervention: a protocol through which two agents communicate. \textbf{(C)} depicts a UI-based intervention, such as controls for interrupting the agent mid-operation.}
    \label{fig:thoughts}
\end{figure}

\subsection{Examples of Design Patterns from Analysis}
\label{a:pattern-examples}

\subsubsection{Visible thoughts, plans, and actions}
\label{a:visible-cot}

\begin{figure}[h]
    \centering
    \includegraphics[width=1.0\textwidth]{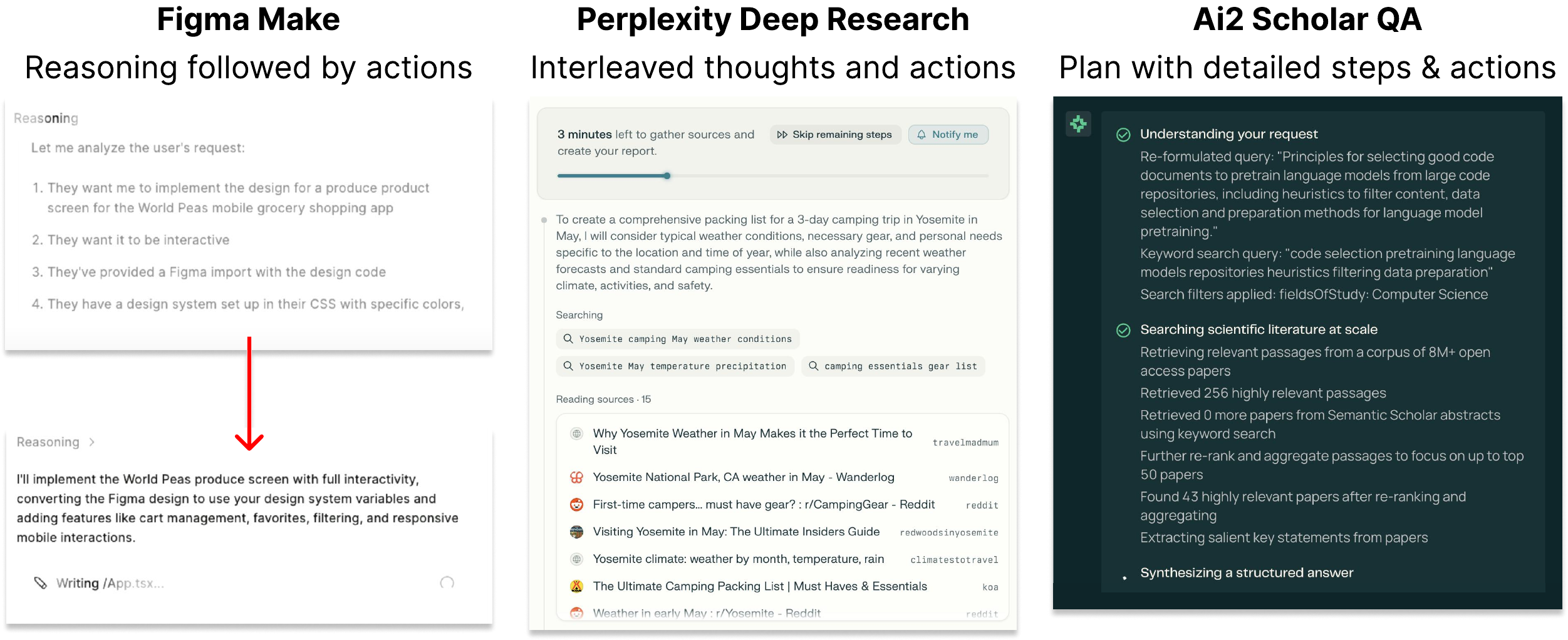}
    \caption{Examples of systems that make agentic reasoning and actions visible to the user, as part of the \textit{visible thoughts, plans, and actions} design pattern.}
    \label{fig:thoughts-cot}
\end{figure}

\begin{figure}[h]
    \centering
    \includegraphics[width=1.0\textwidth]{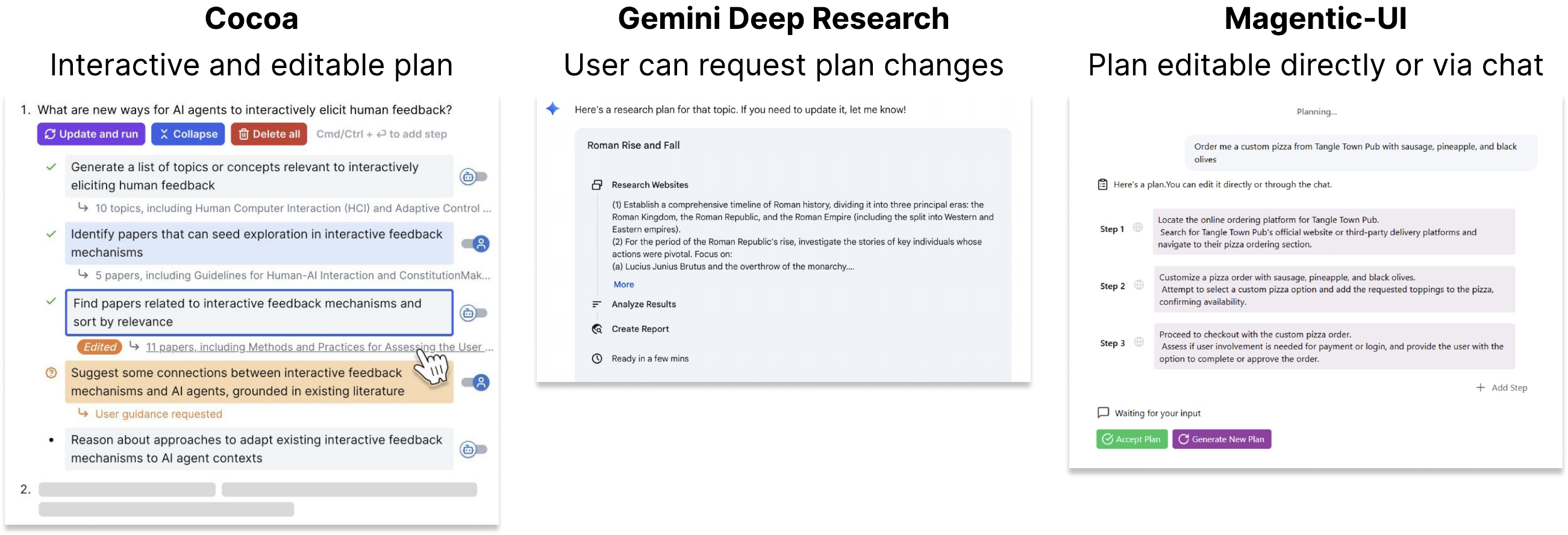}
    \caption{Examples of systems that provide a plan of action and allow the user to edit the plan before execution, as part of the \textit{visible thoughts, plans, and actions} design pattern.}
    \label{fig:thoughts-cot}
\end{figure}

\newpage
\subsubsection{Mechanisms for control transfer}
\label{a:control-transfer}
\begin{figure}[h]
    \centering
    \includegraphics[width=0.9\textwidth]{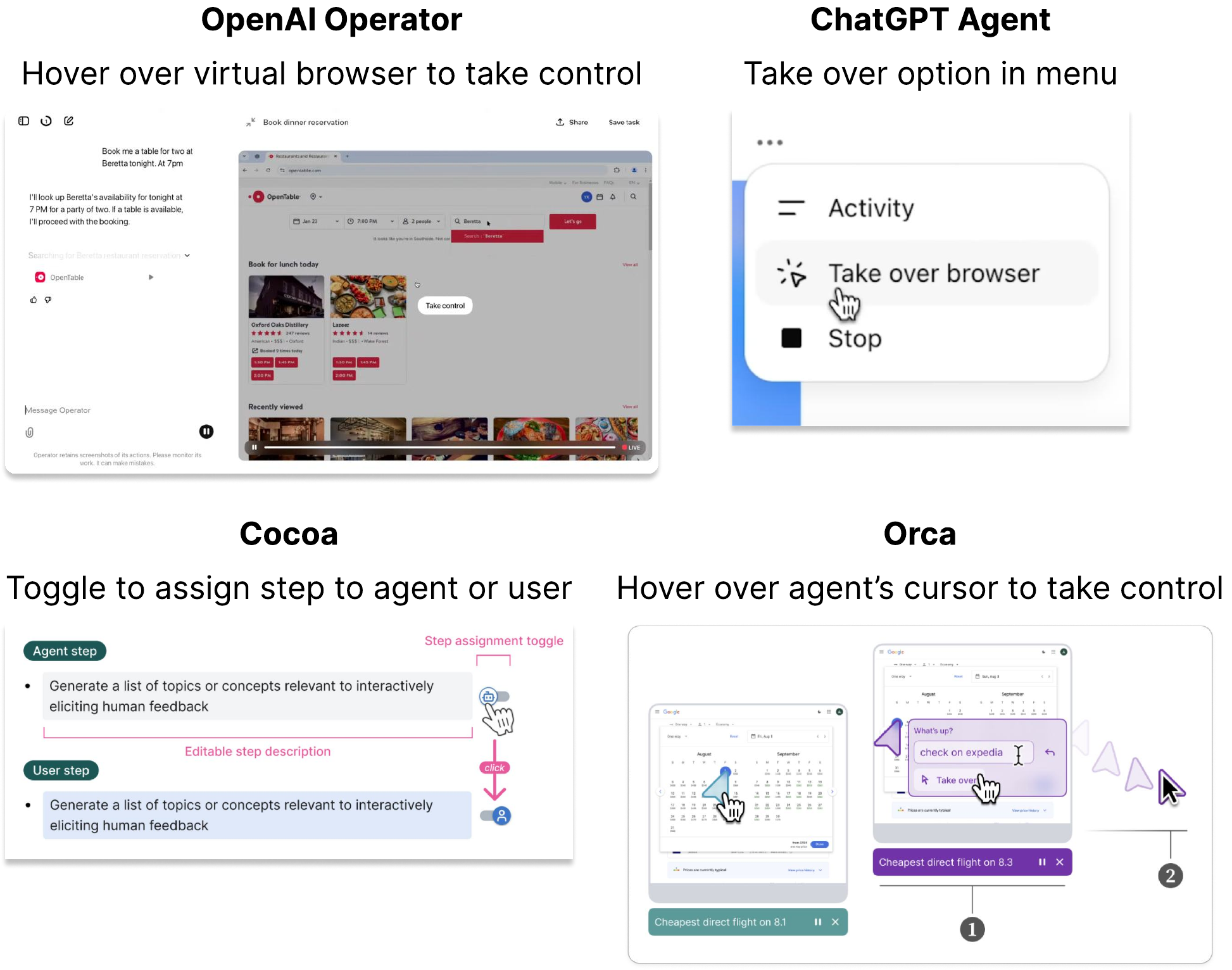}
    \caption{Examples of the \textit{mechanisms for control transfer} design pattern.}
    \label{fig:control}
\end{figure}

\newpage
\subsubsection{Watch mode}
\label{a:watch-mode}
\begin{figure}[h]
    \centering
    \includegraphics[width=0.7\textwidth]{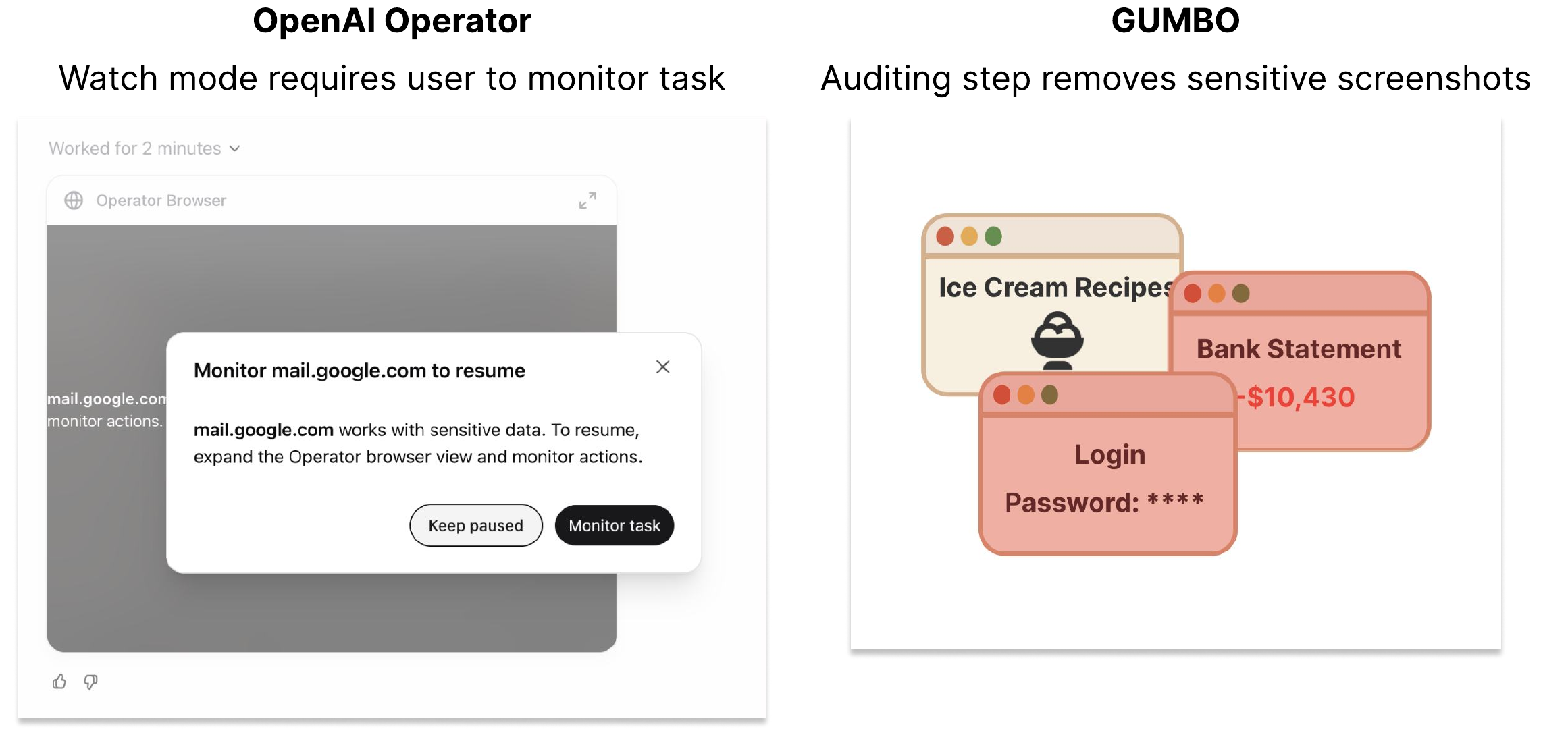}
    \caption{Examples of the \textit{watch mode} design pattern.}
    \label{fig:watch}
\end{figure}

\subsubsection{Customizable approval-seeking conditions}
\label{a:approval-seeking}
\begin{figure}[h]
    \centering
    \includegraphics[width=1.0\textwidth]{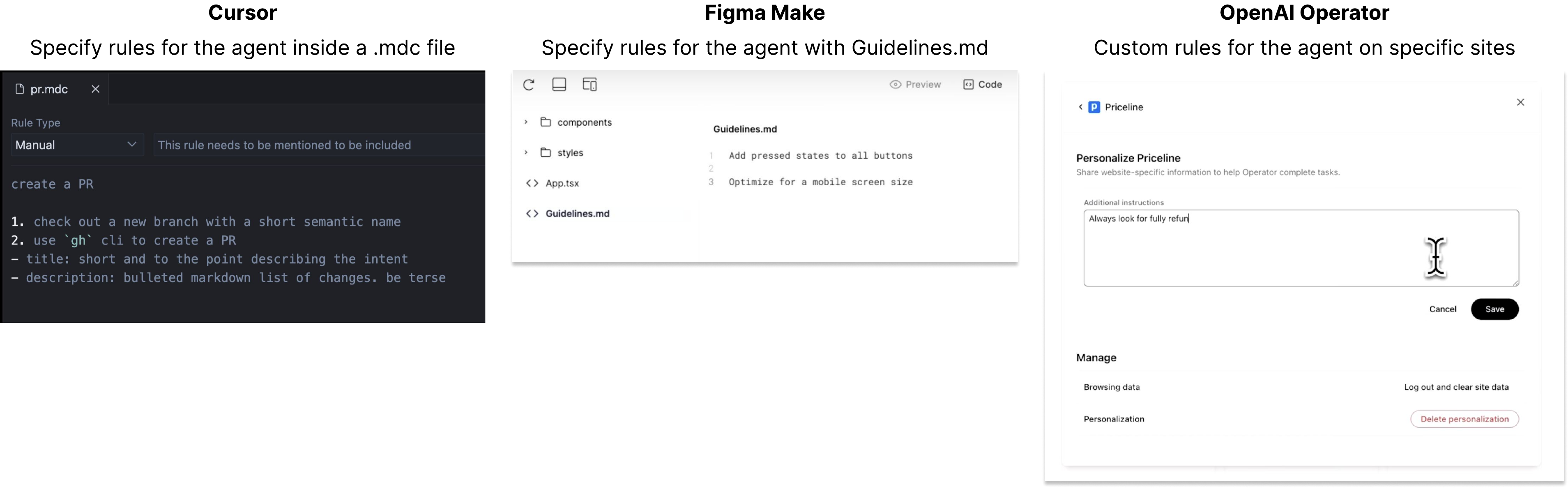}
    \caption{Examples of the \textit{customizable approval-seeking conditions} design pattern.}
    \label{fig:approval}
\end{figure}

\subsubsection{Browsable and editable agent memory}
\label{a:memory}
\begin{figure}[h]
    \centering
    \includegraphics[width=0.7\textwidth]{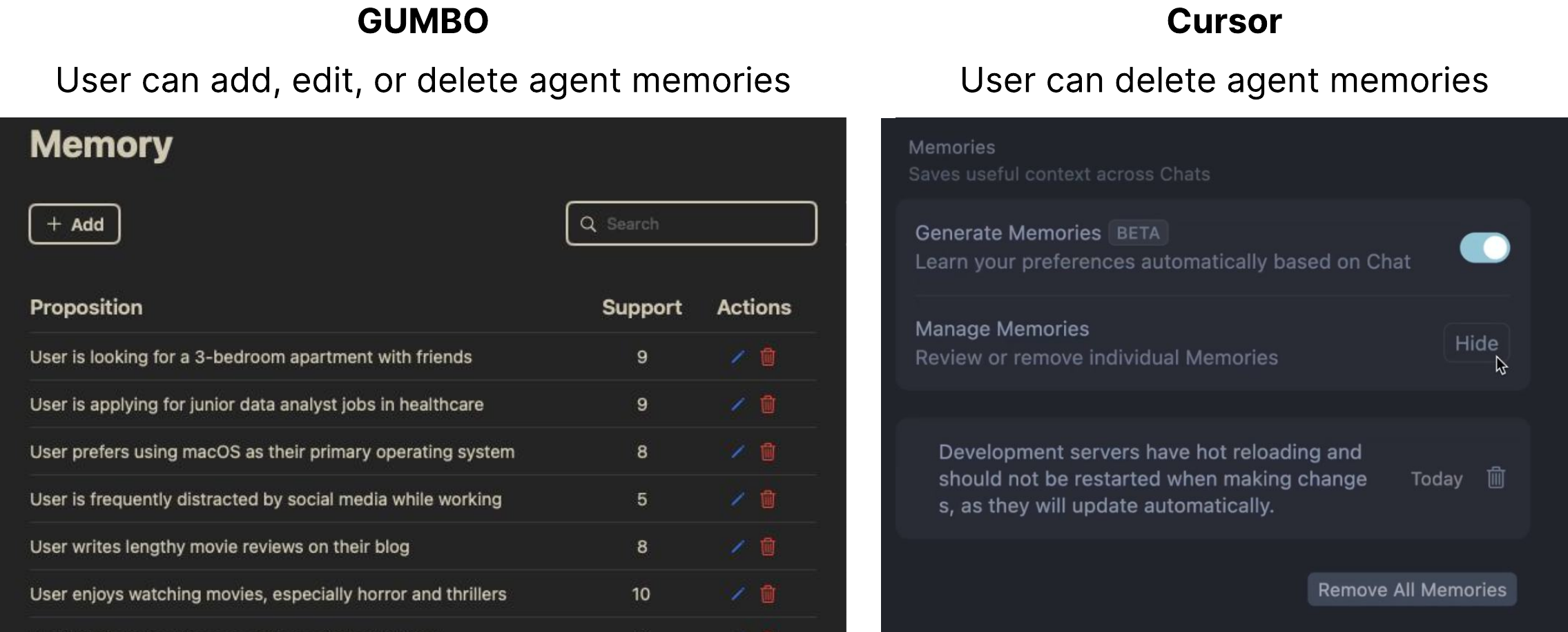}
    \caption{Examples of the \textit{browsable and editable agent memory} design pattern.}
    \label{fig:memory}
\end{figure}

\newpage
\subsubsection{Sandboxes for agents with low-level environmental control}
\label{a:sandbox}
\begin{figure}[h]
    \centering
    \includegraphics[width=1.0\textwidth]{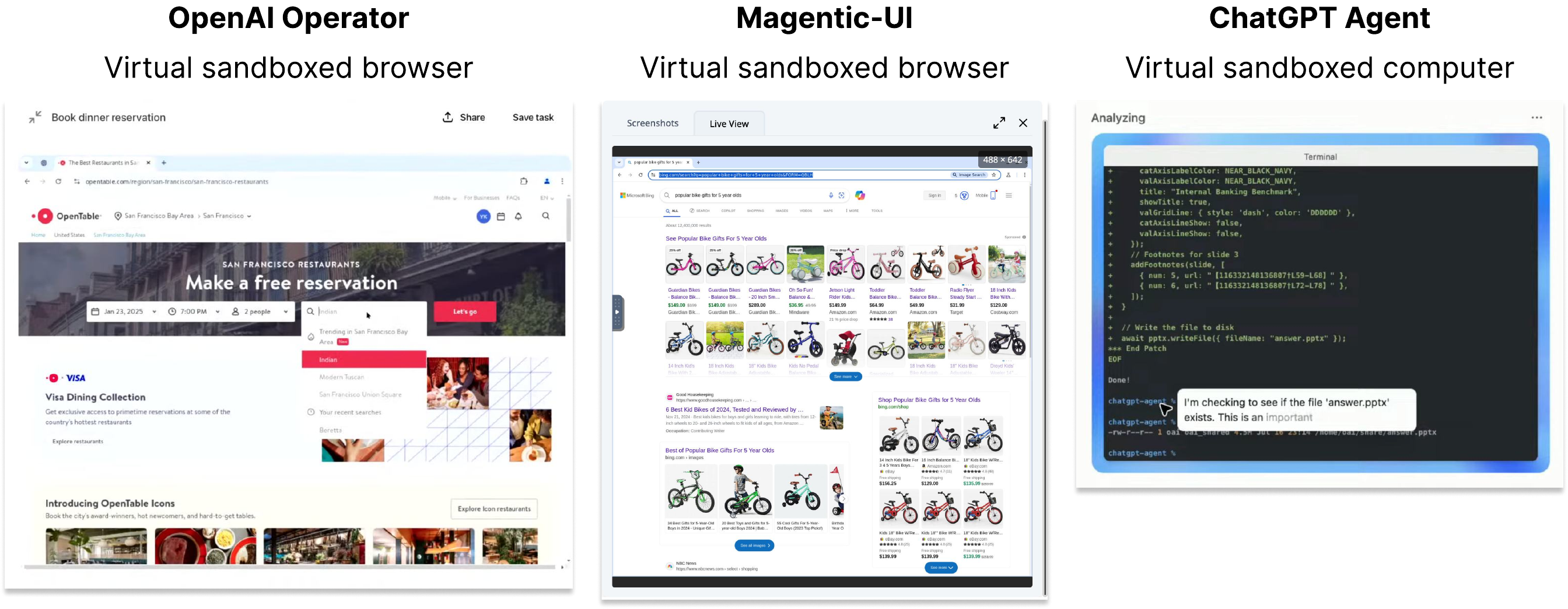}
    \caption{Examples of the \textit{sandboxes} design pattern. Note that one of the systems analyzed have built-in indicators of sandbox status and health.}
    \label{fig:sandbox}
\end{figure}

\subsection{Inclusion Criterial Details}
\label{a:inclusion-criteria}
\begin{enumerate}
    \item \textbf{Is publicly available.} The system is available to use for the public (e.g., not in private beta). Alternatively, the system has detailed documentation of its functionality and interactive components via a paper, blog post, and/or video demo available to the public.
    \item \textbf{Is an interactive software system.} The system affords continuous user interaction through a graphical user interface (GUI) and/or a command line interface (CLI).
    \item \textbf{Operates using multi-step workflows.} The system plans, reasons, and acts over two or more action-taking steps. 
    \item \textbf{Calls tools and/or executes actions.} The system uses external software tools (e.g., APIs) to perform actions that an unscaffolded LLM cannot perform alone.
\end{enumerate}

\subsection{Analysis Method Details}
\label{a:method-details}
The authors read the available papers and watched the available demos, while also testing the system directly where possible. Screenshots were taken throughout the process to document individual elements within each agent UI. The UI element screenshots were collected in a shared FigJam\footnote{FigJam is a collaborative whiteboarding tool: \url{https://www.figma.com/figjam/}.} board, labeled with a short description of its functionality, and clustered based on common functionalites. The UI elements and labels were discussed at weekly meetings. The authors then synthesized the UI elements into six higher-level \textit{interaction design patterns}---general design solutions to issues arising in UIs and UX \cite{design-patterns}---based on functional similarity and repeated use across different systems. 

\newpage
\subsection{Agentic Systems Analyzed}
\label{a:systems}

\begin{table}[ht]
    \centering
    \begin{tabular}{p{3cm}|p{3cm}|p{2cm}|p{5cm}}
    \toprule
         \textbf{Name} & \textbf{Domain} & \textbf{Environment} & \textbf{URL (system or paper)} \\
\midrule
AGDebugger \cite{epperson2025interactive} & Multi-agent systems & Specialized application & \url{https://github.com/microsoft/agdebugger}\\
Ai2 ScholarQA \cite{singh2025ai2} & Scientific research & Web & \url{https://scholarqa.allen.ai/chat}\\
ChatGPT Agent & Computer use & Computer & \url{https://openai.com/index/introducing-chatgpt-agent/}\\
Claude Code & Coding & Terminal & \url{https://www.anthropic.com/claude-code}\\
Cocoa \cite{feng2024cocoa} & Scientific research & Specialized application & \url{https://arxiv.org/abs/2412.10999}\\
CowPilot \cite{huq2025cowpilot} & Browser use & Web & \url{https://arxiv.org/abs/2501.16609}\\
Cove & General productivity & Specialized application & \url{https://cove.ai}\\
Cursor Agent Mode & Coding & IDE & \url{https://docs.cursor.com/en/agent/modes#agent}\\
Figma Make & Design & Specialized application & \url{https://www.figma.com/make/}\\
Gemini Deep Research & General productivity & Web & \url{https://gemini.google/overview/deep-research/?hl=en}\\
GitHub Copilot Agent Mode & Coding & IDE & \url{https://github.com/features/copilot}\\
Gumbo \cite{shaikh2025creating} & Computer use & Computer & \url{https://arxiv.org/abs/2505.10831}\\
Interactive task decomposition \cite{kazemitabaar2024improving} & Data Analysis & Specialized application & \url{https://arxiv.org/abs/2407.02651}\\
Jules & Coding & Specialized application & \url{https://blog.google/technology/google-labs/jules/}\\
Lovable & Coding & Specialized application & \url{https://lovable.dev/}\\
Magentic-UI \cite{mozannar2025magentic} & Browser use & Web & \url{https://microsoft.github.io/magentic-ui/}\\
Manus & General productivity & Web & \url{https://manus.im/}\\
OpenAI Deep Research & General productivity & Web & \url{https://openai.com/index/introducing-deep-research/}\\
OpenAI Operator & Browser use & Web & \url{https://operator.chatgpt.com/}\\
Orca \cite{jiang2025orca} & Browser use & Web & \url{https://orca.jiang.pl/}\\
Perplexity Deep Research & General productivity & Web & \url{https://www.perplexity.ai/hub/blog/introducing-perplexity-deep-research}\\
Proactive programming assistant \cite{chen2025need} & Coding & IDE & \url{https://arxiv.org/abs/2410.04596}\\
         
    \bottomrule
    \end{tabular}
    \vspace{0.1cm}
    \caption{List of agentic systems analyzed. We describe our inclusion criteria in Section \ref{s:method}. Citations are included for systems with academic papers.}
    \label{t:systems}
\end{table}